\documentstyle{camera}

\newcommand{\AmS}{{\protect\the\textfont2
  A\kern-.1667em\lower.5ex\hbox{M}\kern-.125emS}}

\newcommand{\beq}{\begin{equation}}
\newcommand{\eeq}{\end{equation}}
\newcommand{\bea}{\begin{eqnarray}}
\newcommand{\eea}{\end{eqnarray}}

\def\dm2{\Delta m^2}
\def\sq2{sin^2(2\Theta)}

\input epsf

\begin{document}

%
\title{ASTROPARTICLE PHYSICS WITH AMS02}

%
\author{BEHCET ALPAT}

%
\organization{(on behalf of the AMS-02 Collaboration)\\
Istituto Nazionale di Fisica Nucleare, Sezione di Perugia\\
Via Alessandro Pascoli 1, 06123, Perugia, Italy}

\maketitle

\begin{abstract}
 The Alpha Magnetic Spectrometer (AMS02) experiment will be
installed in 2009 on the International Space Station (ISS) for an
operational period of at least three years. The purpose of AMS02
experiment is to perform accurate, high statistics, long duration
measurements in space of charged cosmic rays in rigidity range from
1 GV to 3 TV and of high energy photons up to few hundred of GeV. In
this work we will discuss the experimental details and the physics
capabilities of AMS02 on ISS.
\end{abstract}
\vspace{1.0cm}

\section{Introduction}
In June 1998 a reduced version of the Alpha Magnetic Spectrometer
(AMS01) experiment has successfully flown for 10 days on Space
Shuttle Discovery (STS-91). This mission has provided valuable
information on detector performance in actual space conditions and
interesting cosmic ray data. The detector layout, performance and
the physics results of AMS01 during STS-91 mission are described in
detail elsewhere (Aguilar et al., 2002).


The AMS02 is a multi purpose detector aiming to study, with
unprecedented sensitivity, cosmic antimatter, the indirect search of
dark matter constituents and perform high accuracy and high
statistics charged cosmic ray spectra up to TeV region. AMS02 will
also measure the high energy gamma rays up to few hundred GeV with
good source pointing capability.

\section{AMS02 Experiment and Detector Performance}
In order to achieve physics goals, the detector requirements include
large acceptance ($\sim$0.5 $m^2\cdot sr$), accurate particle
identification and  rigidity ($p\over Z$), energy, charge
measurements, as well as good e/p separation and the system
redundancy. Fig.1 shows an exploded view of AMS02 experiment
together with tracker rigidity resolution and dE/dX performance for
nuclei charge determination.

\begin{figure}[t!]
      \vspace{10.5truecm}
\includegraphics{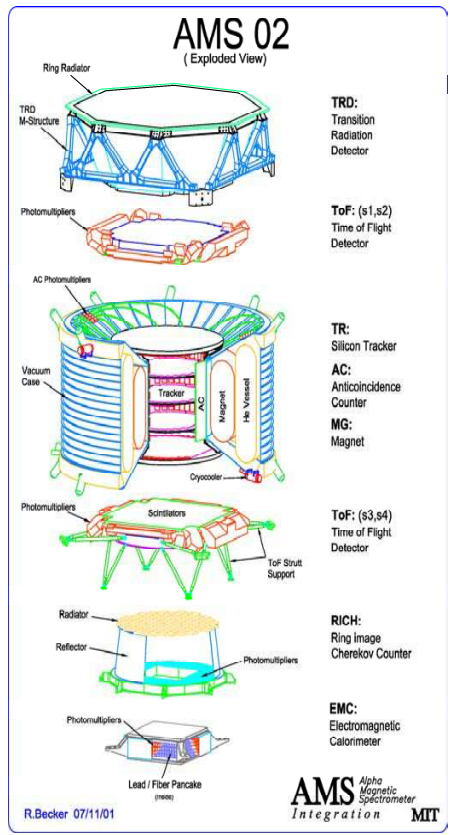}
\includegraphics{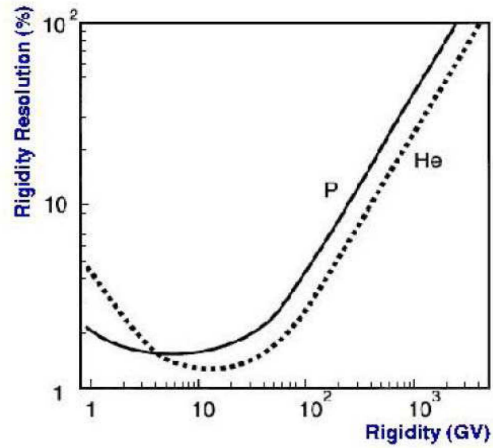}
\includegraphics{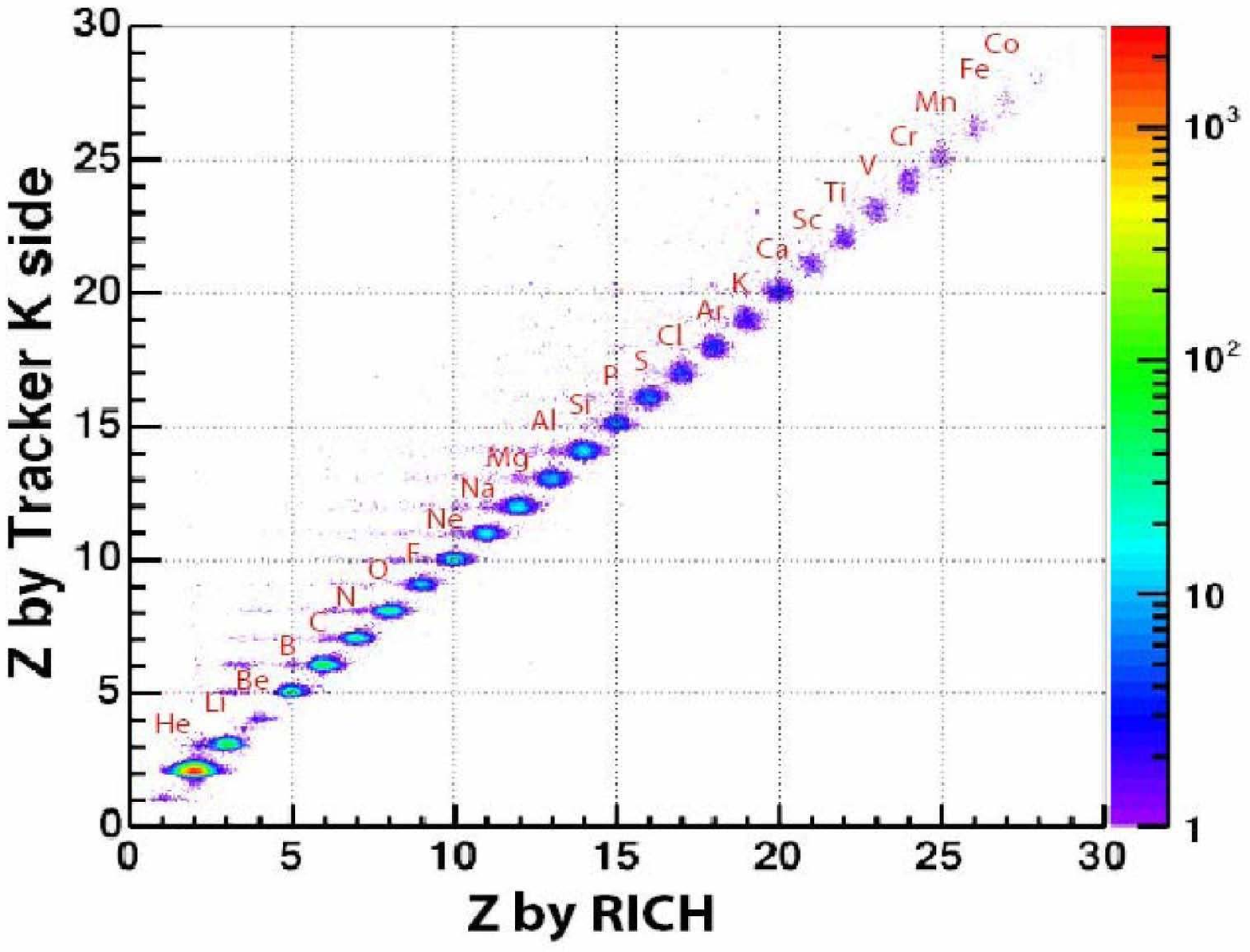}
      \caption[h]{The AMS02 exploded view (Left). The silicon
      tracker rigidity resolution for p and He(upper right). The nuclei separation with dE/dX
      measured both by silicon tracker and ring imaging
      Cherenkov detector (lower right).}
     \label{fig1}
    \end{figure}

In the following the AMS02 sub-detectors will be briefly described
from the upper part of the experiment.
\begin{itemize}
  \item The Transition Radiation Detector (TRD) is designed to suppress the proton
signal with a rejection factor against positrons of $10^3$-$10^2$ in
the energy range from 10-300 GeV. The TRD consists in 20 layers of 6
mm diameter straw tubes alternating with fleece radiators. The
strawtubes are filled with a 80\%/20\% mixture of Xe/ $CO_2$ at 1
bar. Combined with the Electromagnetic CALorimeter (ECAL)
performance, an overall $e^+$/p rejection factor of $\sim 10^6$ at
90 \% of $e^+$ efficiency will be achieved.

\item The Time of Flight (ToF) system consists of four planes of  plastic scintillator
placed at both ends of superconducting magnet. Lightguides were
designed to minimize the angles between local magnetic field and PMT
axis to optimize their response. The ToF is designed to provide fast
trigger to the experiment, measurement of time of flight of the
particles traversing the detector with up/down separation and  dE/dX
measurements. The $\Delta{\beta}/\beta$ is 3\% for protons and the
estimation on absolute particle charge can be done up to Z=20. The
beam tested overall time resolution is 160 ps for protons and better
for heavier cosmic ray nuclei.

\item The SuperConducting Magnet (SCM) bore has a diameter of 1.1 m in
which the silicon tracker is mounted. The SCM  consists of two
dipole and two sets of smaller racetrack coils with a total cold
mass of about 2300 kg. The racetrack coils are designed to increase
the overall dipole field, to minimize the stray dipole field outside
the magnet (max stray field at a radius of 3 m is 4 mT). It has a
bending power of $B\cdot L^2\sim$ 0.8 $Tm^2$.
 All coils are wounded with high purity
aluminum-stabilized niobium-titanium conductor. The magnet will be
operated at a temperature of 1.8 K and cooled by 2500 l of
superfluid helium, and it should be operational for three years
without refilling.

\item
The Silicon Tracker of AMS02 is designed to perform high precision
rigidity measurements, and to determine the sign and the absolute
value of particle charge. The ST consists of 8 thin layers of
double-sided, 300 $\mu$m thick, silicon microstrip detectors of
lengths up to 60 cm. The ST has 0.3 \% of total radiation length and
6.7 $m^2$ of active surface. The readout electronics is based on low
noise, low power, high dynamic range VA64-HDR9 VLSI, circuitry. The
charge determination capability combined with that of Ring Imaging
CHerenkov (RICH) detector is given in Fig.1(Lower Right). In Fig.
1(Upper Right) an estimate of the AMS02 proton rigidity resolution
($>$5 hit track) for protons and helium is shown.

The tracker cooling system bases on two-phase mechanically pumped
closed circuit in which cooling fluid ($CO_2$) runs by capillary
forces. The fluid enter the tracker just under the boiling point to
collect the heat and then outgoing fluid/vapor mixture is cooled on
thermal radiator panels. The system dissipates 150 W.

\item
In AMS02, Ring Imaging CHerenkov Detector (RICH) provides additional
velocity measurement with $\Delta{\beta}/\beta$ of 0.1\% up to Z=26
and the isotopic separation is covered in the energy range from 0.5
GeV/n to 10 GeV/n for A $\le$ 10. The RICH is a proximity focusing
device with a dual solid radiator at the top, an expansion volume at
the center and a matrix of multipixelized photon readout cells at
the bottom.

\item
Below RICH, the AMS02 includes a fine grained sampling
Electromagnetic CALorimeter (ECAL, 16 $X_0$ and 18 samplings) for
3-D imaging of shower development hence discrimination between
hadronic and electromagnetic cascades. ECAL is a sampling device
with a lead-scintillating fibers structure with 9 superlayers (X and
Y views) each containing 11 grooved Pb foils interleaved with 10
scintillator fiber layers glued with epoxy resin. The design goal
for ECAL is to provide precise (dE/E $<$ 5 \%) e$^-$,e$^+$ and
$\gamma$ spectrum from 1 GeV to  1 TeV and good e/p discrimination
($O(10^3)$ for $<$500 GeV). The TRD+ECAL and ST combined e/p
separation is $>10^6$. Moreover, for gamma ray studies, ECAL acts as
independent photon detector with an angular resolution of $\sim$
1$^\circ$.

\item
The star tracker system (AMICA) will give a precise measurement of
the AMS02 observing direction with a few arc-sec accuracy.
\end{itemize}

In ASM02 there are a total of $\sim$300,000 electronic channels
delivering about 7 Gbit/s of raw data. The predicted total trigger
rate varies from 200 to 2000 Hz. The DAQ electronics reduces the
event size, through proper filtering, to the allocated downlink data
rate of 2 Mbit/s.

All electronics and mechanical parts of AMS02 are tested for
operation in vacuum, EMI/EMC compatibility, vibration and thermal
cycles. The effect of total ionization dose (up to 6 Gy/year) on all
critical components is extensively tested. The AMS02 weighs about 7
tons and has a power consumption of about 3 kW.
\section{AMS02 Physics}
\subsection{Antimatter Search}
 The excess of baryonic
matter over antimatter is characterized by the observed ratio
$\eta=(n_{B}-n_{\bar B})/n_\gamma \cong 10^{-10}$.  Over the time to
evolve the initially symmetric universe  into today's matter
dominated one (baryogenesis), according to Andrei Sakharov (A.
Sakharov, 1967), three principles should be fulfilled: non
conservation of baryonic charge, breaking of C and CP invariance and
the deviation from thermal equilibrium. Though there are several
theories (Soni, 1997) suggest that the quantum effects allow
universe to tunnel between vacua with different baryon number (B)
values and this tunneling may occur at future supercollider energies
energies ($>$10 TeV, the sphaleron mass), at present there is no
experimental evidence that B is violated. Moreover the Belle and
Babar experiments has looked into the violation parameter
sin2$\beta$ (=sin2$\phi_1$) using $B^0\rightarrow J/\Psi K^0$
channel (Daniel R. Marlow, 2003; S. Noguchi, 2003). Two independent
experiments employ different detectors and analysis techniques but
nonetheless yield results consistent with one another and with the
Standard Model expectations based on measurements of other CKM
matrix parameters. The results are sin2$\beta =0.75\pm 0.09\pm 0.04$
(BaBar 56 $fb^{-1})$ and sin2$\phi_1 =0.82\pm 0.12\pm 0.05$ (Belle
42 $fb^{-1})$.

There exist different inhomogeneous baryogenesis models, mostly
based on assumptions that different sign of C(CP) breaking in
different space points and moderate blow-up of regions with a
definite sign of charge symmetry breaking. They predict matter and
antimatter regions  (Dolgov, 2002; Kirilova 2002) with antimatter
structures like antigalactic clusters, antigalaxies situated between
clusters of galaxies and antistar globular clusters providing
signatures in $^3{\bar{He}}$ and $^4{\bar{He}}$ spectrums. Present
antihelium/helium ratio limits and expected AMS02 sensitivity for
three years on ISS are given in Fig.2(Left).

\begin{figure}[t!]
      \vspace{6truecm}
\includegraphics{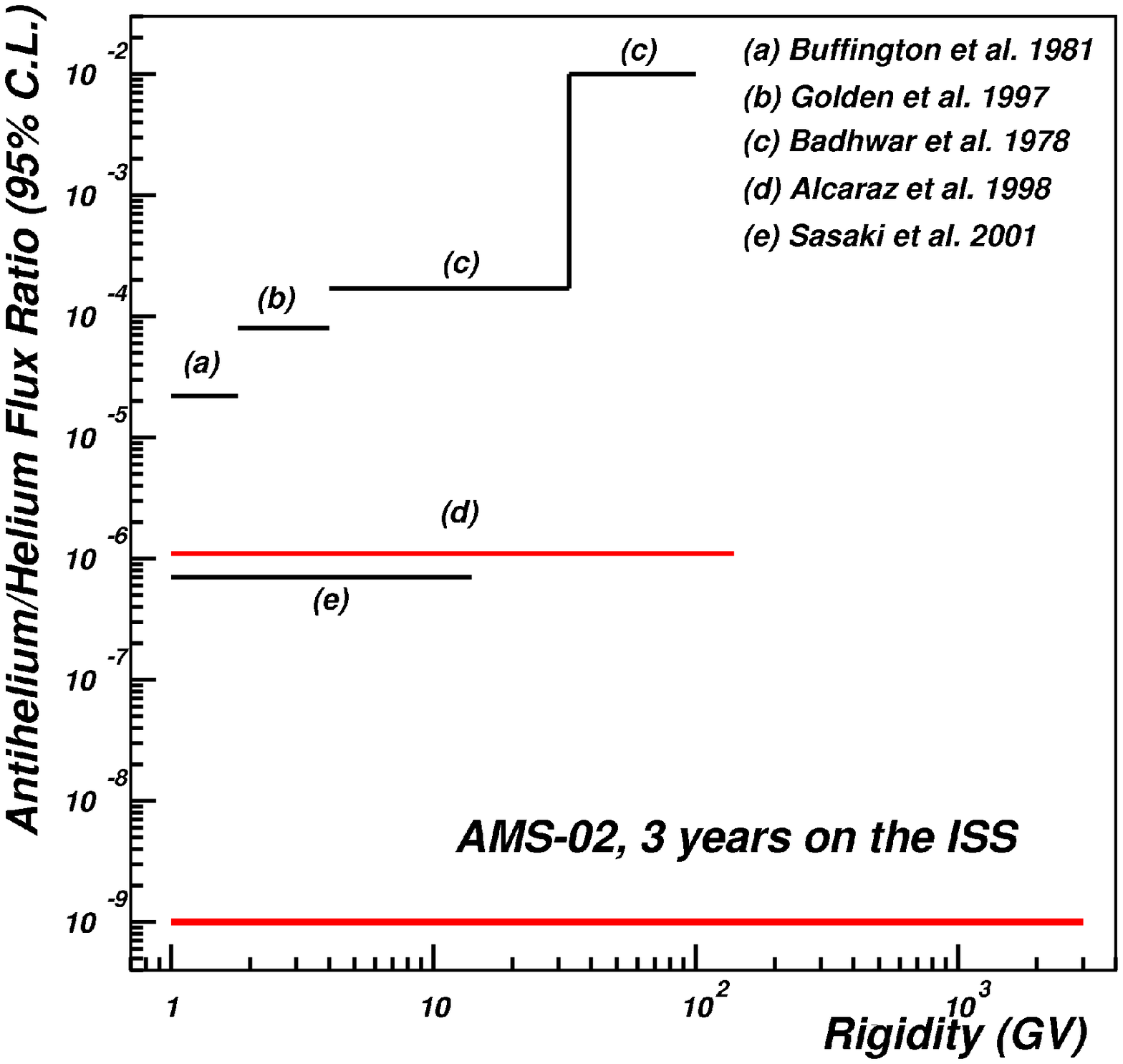}
\includegraphics{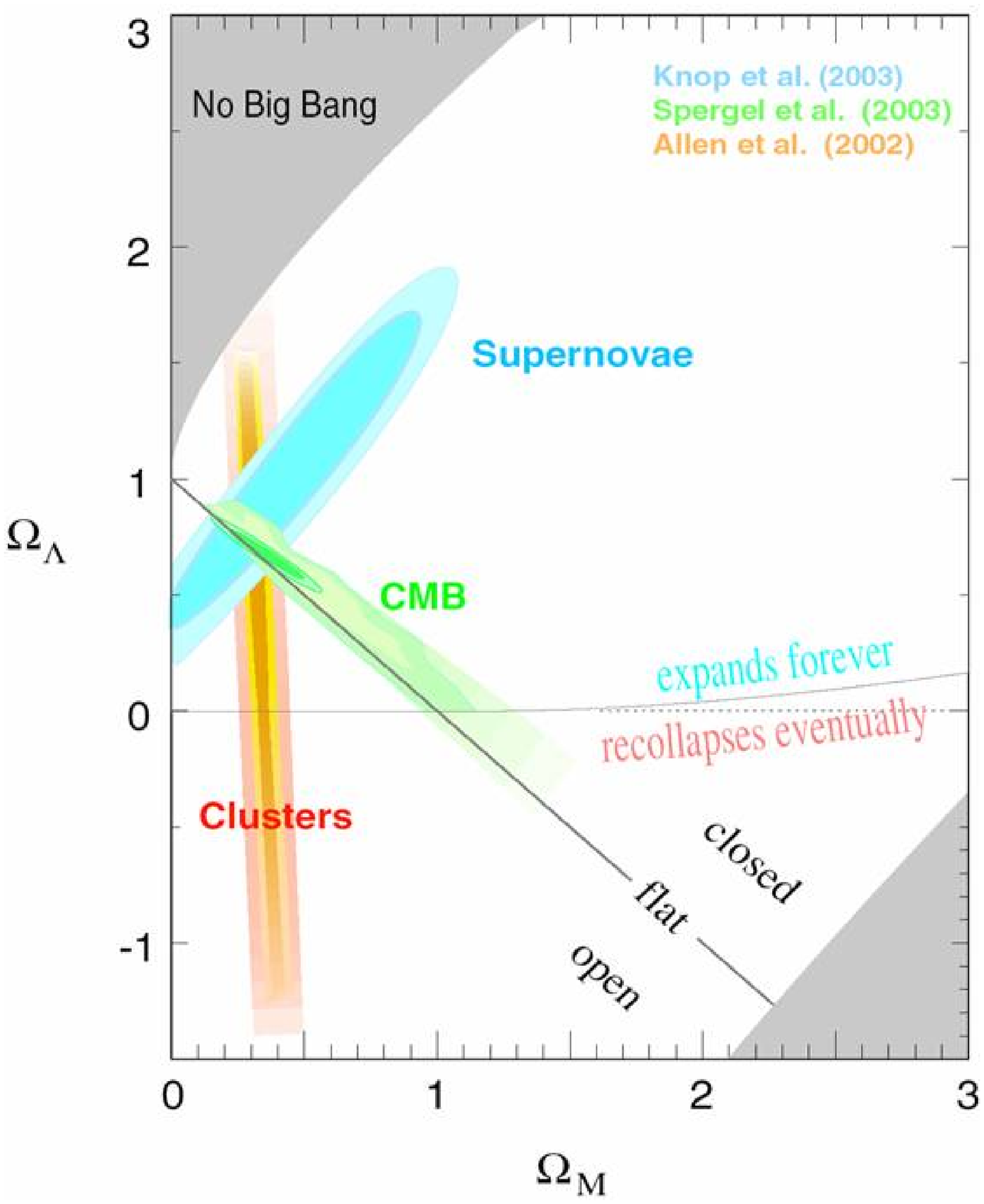}
 \caption[h]{(Left):The ${\bar{He}\over He}$ flux ratio as a
function of rigidity for present data and the expected AMS02 limit
after three years on ISS. (Right): $\Omega_{\Lambda}$-$\Omega_{m}$
plane; all data is converging for a flat and expanding Universe.}
     \label{fig2}
    \end{figure}

\subsection{Dark Matter}
At all scales (galaxies, clusters, superclusters...) the visible
mass is not sufficient to explain the observed dynamical effects. As
it can be seen from the Fig.2(Right) that in the
$\Omega_{\Lambda}$-$\Omega_{m}$ plane all the different
measurements; SNIA brightness observations, Cosmic Microwave
Background (CMB) anisotropies and optical measurements of clusters
of galaxies, converge to a unique point. The universe is flat and
will expand forever (Aldering, G., et al.,2004). In this picture
universe is made of $\sim$ 72\% dark energy, $\sim$23\%
(non-baryonic) darkmatter, $\sim$4\% baryons and $\sim$0.5\%
neutrinos. Dark (negative potential) energy, permeating everywhere
causes the increase on expansion speed and the gravity is not enough
to hold constant the recession velocity.

The weakly interacting massive particles (WIMPs), postulated in
minimal supersymmetric standard model (MSSM) and in other R-parity
conserving supersymmetric models, are particularly attractive to
explain dark matter's nature (Ellis \& Ferstl \& Olive, 2002). In
this framework the lightest supersymmetric particle (LSP), stable
neutralino, $\chi$ , a neutral scalar boson being also its own
antiparticle, is the most quoted candidate. Indirect signals may be
produced by annihilation of neutralinos inside the celestial bodies
where  $\chi$'s have been captured and accumulated. The signals then
emerge from $\chi\chi\rightarrow W^+,W^-,hadrons,b\bar{b}\rightarrow
\gamma\gamma,e^+,\bar{p},\bar{D}$ or
$\chi\chi\rightarrow\gamma\gamma, Z_0\gamma$.

\begin{figure}[t!]
      \vspace{11truecm} \includegraphics{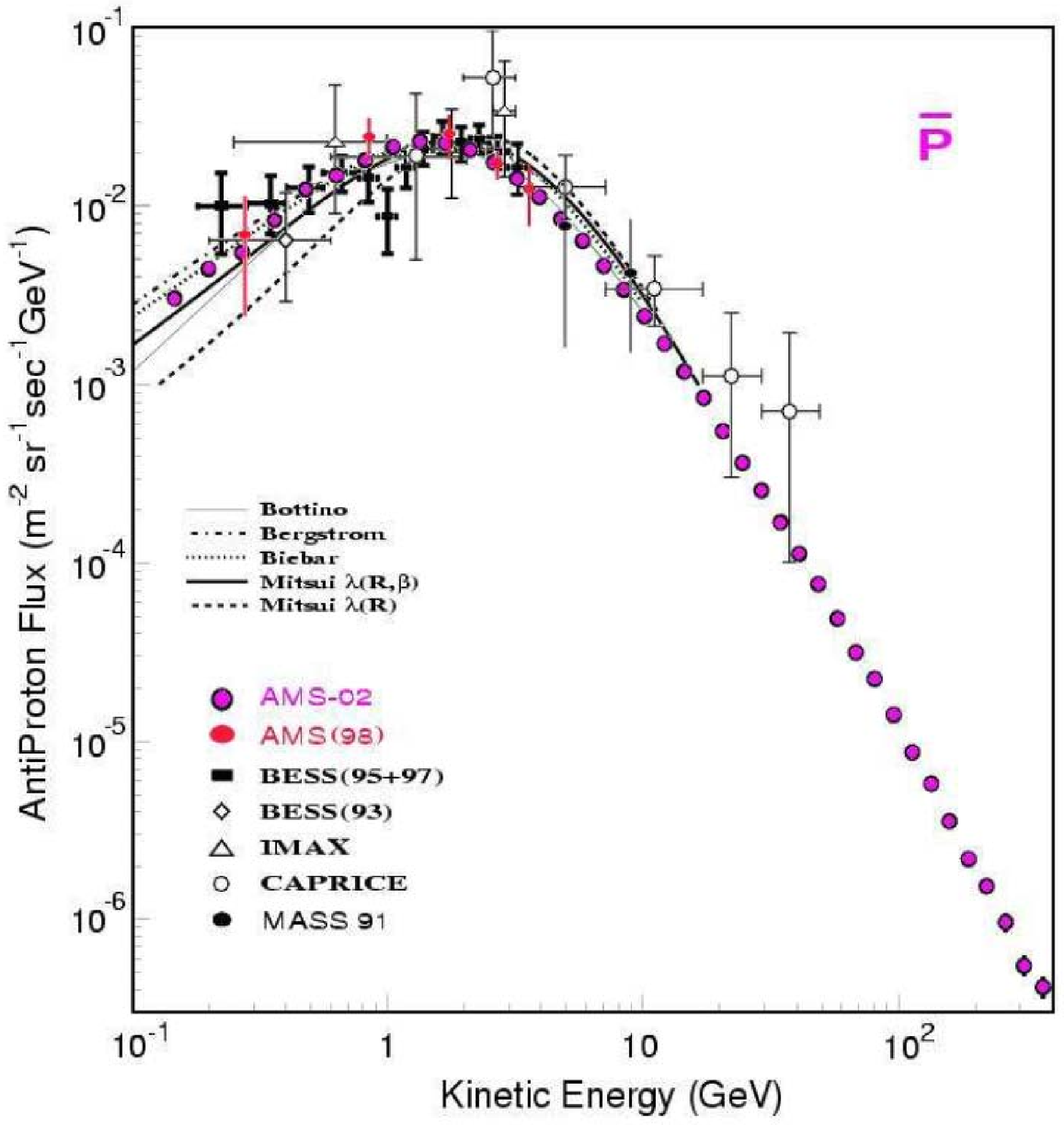}
\includegraphics{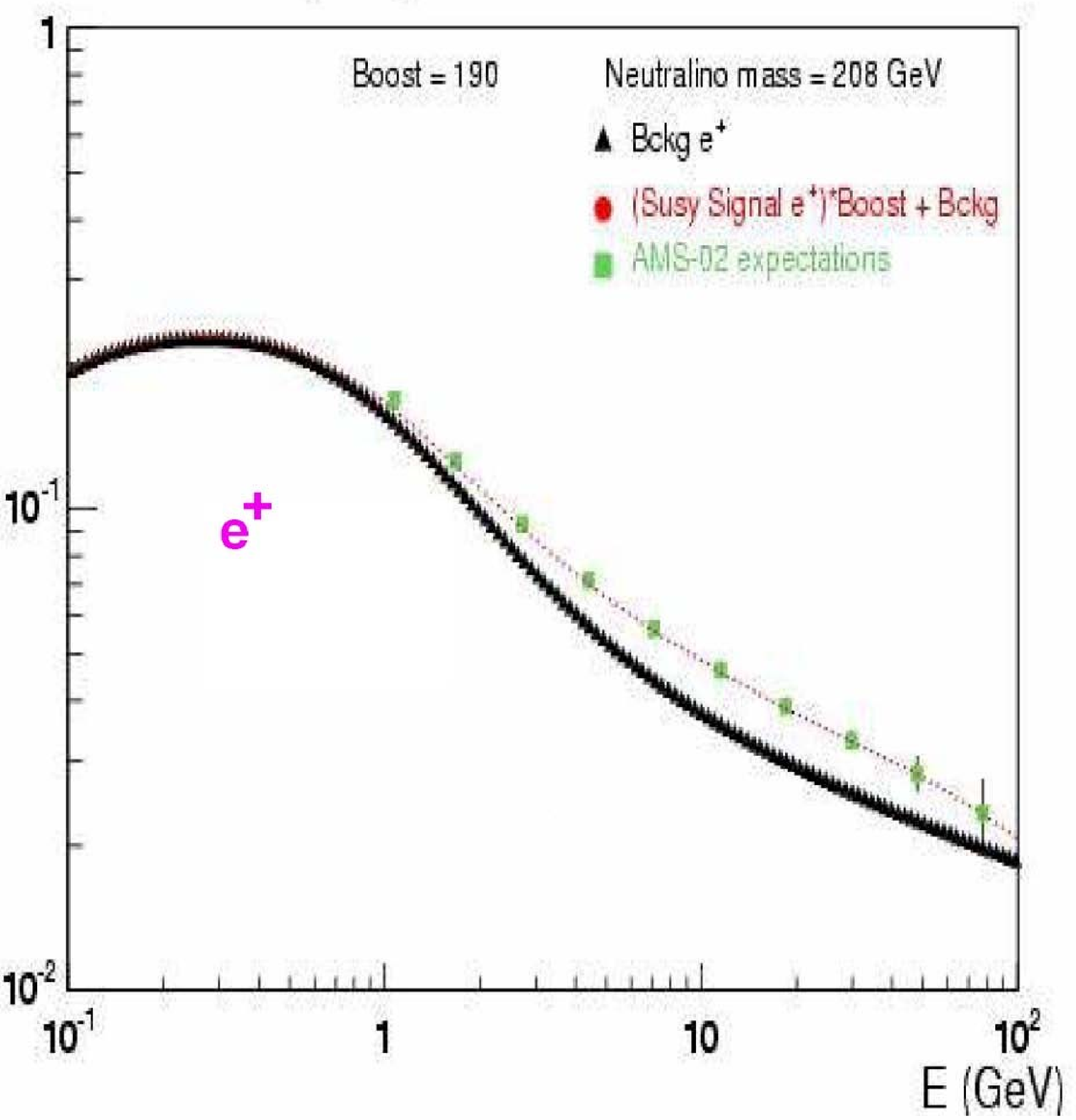}
\includegraphics{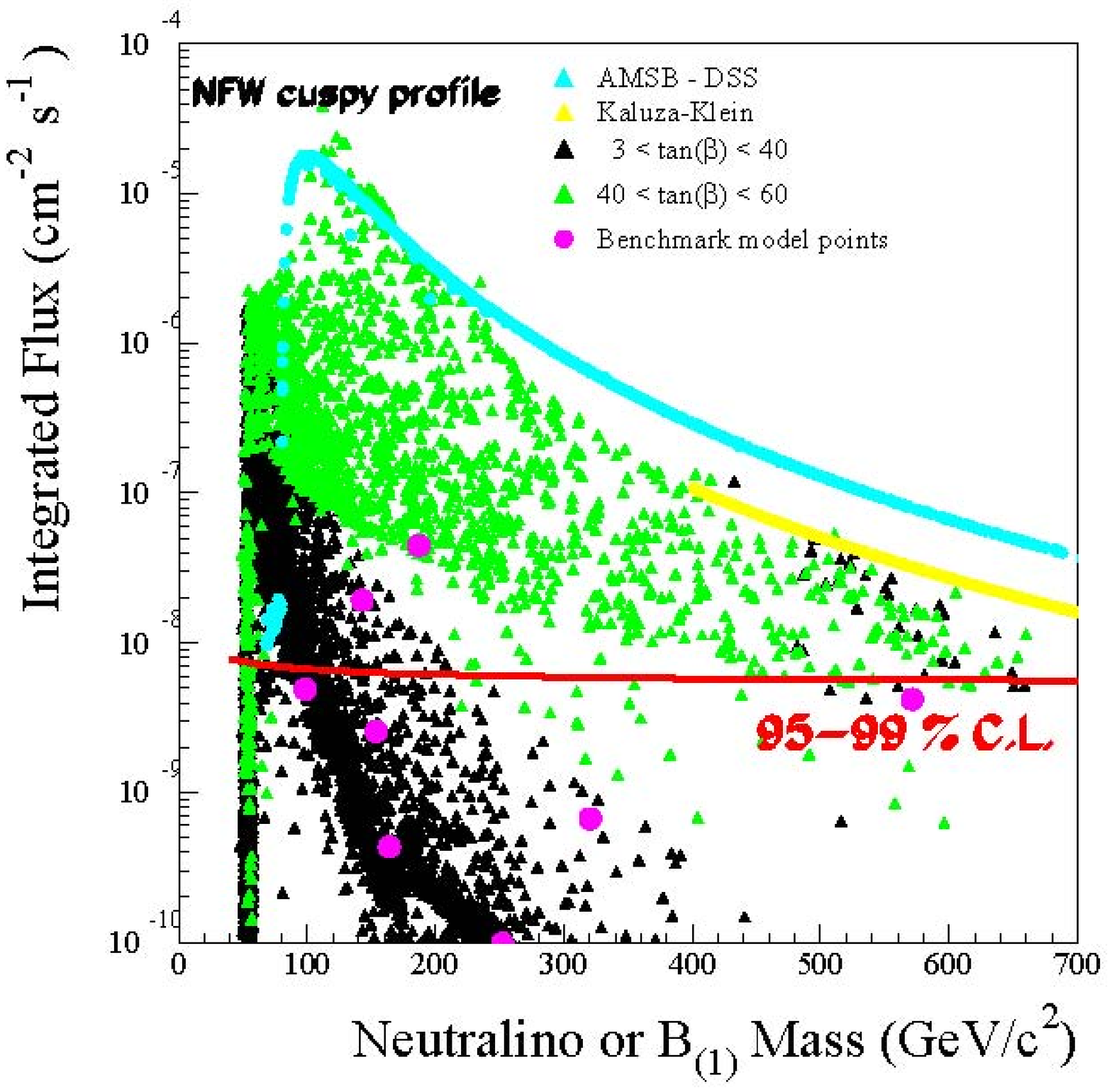}
\includegraphics{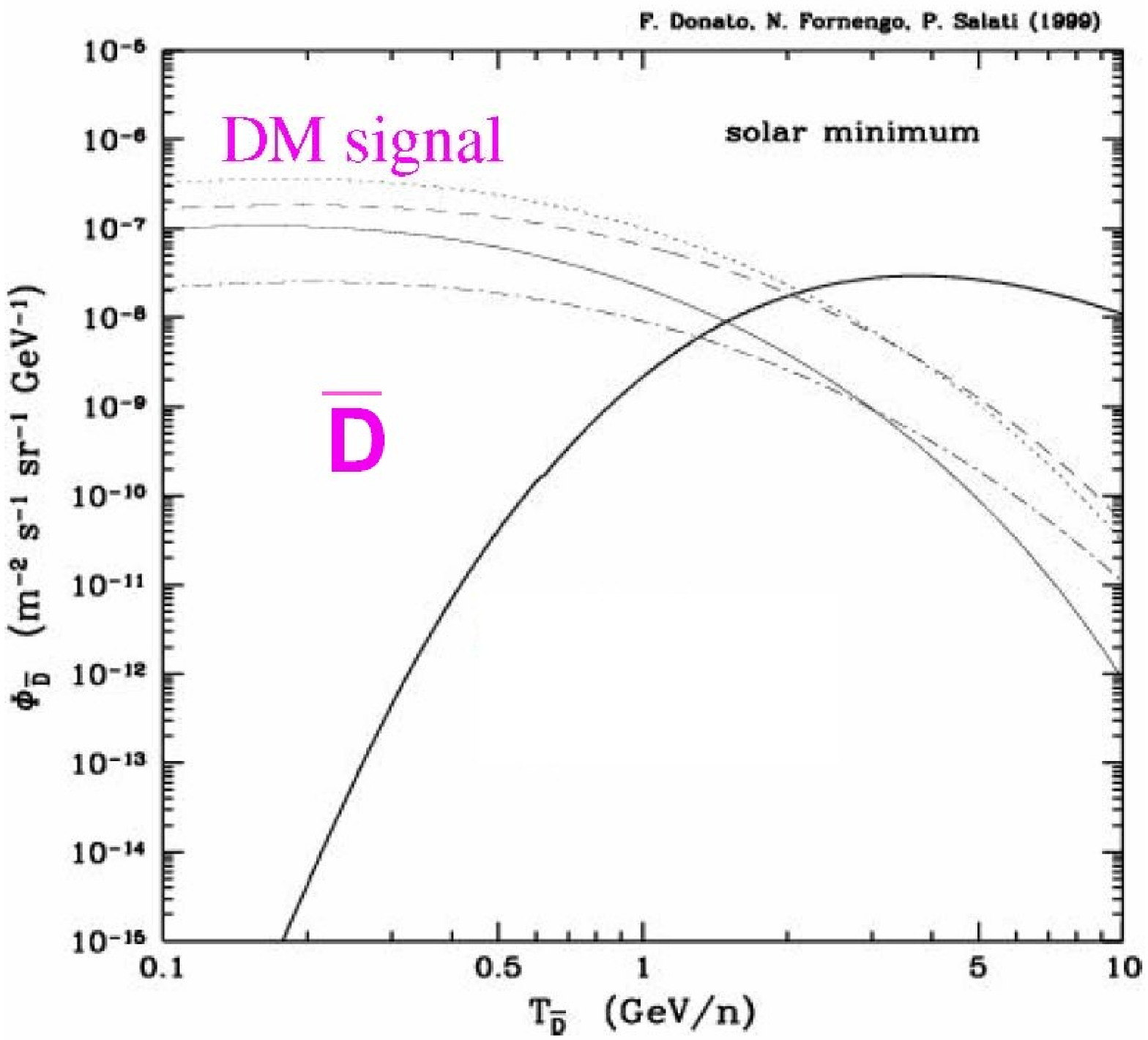}
      \caption[h]{For different neutralino annihilation products detectable by AMS02.
      (UL): $\bar{p}$ flux for present data and expected AMS02 accuracy.
      (LL): integrated $\gamma$ flux from galactic center as a function of $m_{\chi}$ for a
      cuspy NFW dark matter halo profile (Jacholkowska et al., 2005). (UR):
positron excess and AMS02 signal for $m_{\chi}$=208GeV. (LR):
$\bar{D}$ signal (dashed lines) for four different neutralino
annihilation compositions and the background (solid line) at solar
minimum. }
     \label{fig3}
         \end{figure}

The AMS02 will be unique experiment detecting all annihilation
products;
\begin{itemize}
  \item positrons will be measured in (1-300 GeV) with a mean acceptance of 0.045 $m^2.sr$
and a  combined  $e^+/p$ rejection factor of $\sim 10^5$.;
  \item gamma rays will be detected in conversion mode by
  measuring $e^+e^-$ in the tracker and direct measurement in ECAL with no associated charged track activity;
  \item antiprotons will be measured in 0.5-200 GeV with a mean acceptance of 0.03 $m^2\cdot sr$
and a  combined proton rejection factor of $\sim 10^6$;
  \item antideuteron production from proton-proton collisions is a rare
process and it may be less rare in neutralinos annihilation. The
present D/p ratio is about 10$^{-5}$.
\end{itemize}

The Fig.3 shows on upper left, the AMS02 model expectation for high
accuracy antiproton flux compared with present data, on upper right,
deBoer scenario (de Boer, 2004)  with boost factor tuned to match
HEAT+EGRET excess and AMS02 expectation, on lower left, integrated
$\gamma$ ray flux from neutralino annihilation as a function of
$m_{\chi}$. The considered models were $m$sugra and Klaze-Klein
Universal Extra dimensions (Jacholkowska et al., 2005). The results
of the simulations in the framework of $m$sugra model, show that
with a cuspy dark matter halo profile or a clumpy halo, the
annihilation $\gamma$ ray signal would be detectable by AMS02 up to
1 TeV. In Fig. 3, lower right, the (Donato \& Fornengo \& Salati,
2000) model
 for $\chi$ annihilations at the
galactic halo. At lower energies the $\bar{D}$ the signal is well
above background (solid line) at solar minimum.

\section{Cosmic Ray Astrophysics}
The primary and secondary cosmic ray measurements are essential to
determine the backgrounds for weak signal searches. The present
uncertainty on these fluxes is the main contributor for systematic
errors on atmospheric neutrino oscillation calculations. In Fig 4
(Upper Left and Upper Right) are shown the present status for proton
and helium fluxes and the expected AMS02 performance for after few
hours of data collection.

Moreover, the study of relative abundances of elements and isotopes
yields to a better understanding of origin, propagation,
acceleration and confinement time of cosmic rays in our galaxy
(Strong \& Moskalenko, 2001).

The AMS02 mass resolution and charge determination capabilities and
superconducting magnet provides very high sensitivity to determine
the primary, secondary fluxes and heavy nuclei up to iron and for
energies to few TeV.

The AMS02 will be able to measure the particle fluxes with high
accuracy to Z=25 in the energy range 0.1 GeV/n - 1 TeV/n. In Fig.4
(Lower Left) the expected B/C ratio accuracy for 6 months of AMS02
data collection is shown together with model expectation (solid
line) and most recent measurements.  The $^{10}$Be/$^9$Be flux ratio
with $^{10}$Be being lightest unstable isotope with half-life
comparable to the galactic confinement time of cosmic rays, will
provide important hints on galactic halo height and on residence
times of cosmic rays in our galaxy. In Fig 4. (Lower Right) shows
high statistics measurement of $^{10}$Be/$^9$Be flux ratio after one
year.

\begin{figure}[t!]
      \vspace{10truecm}
\includegraphics{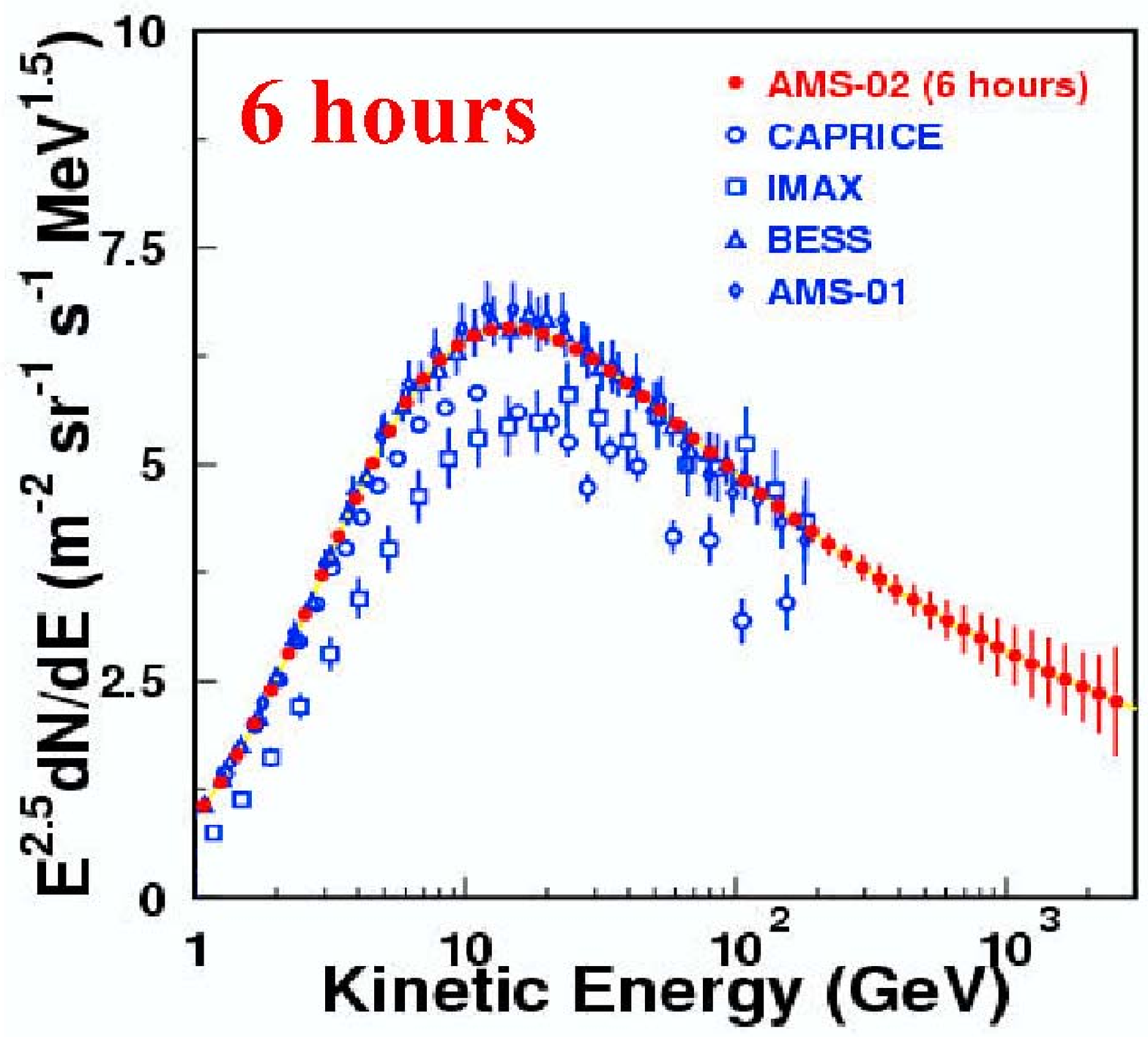}\includegraphics{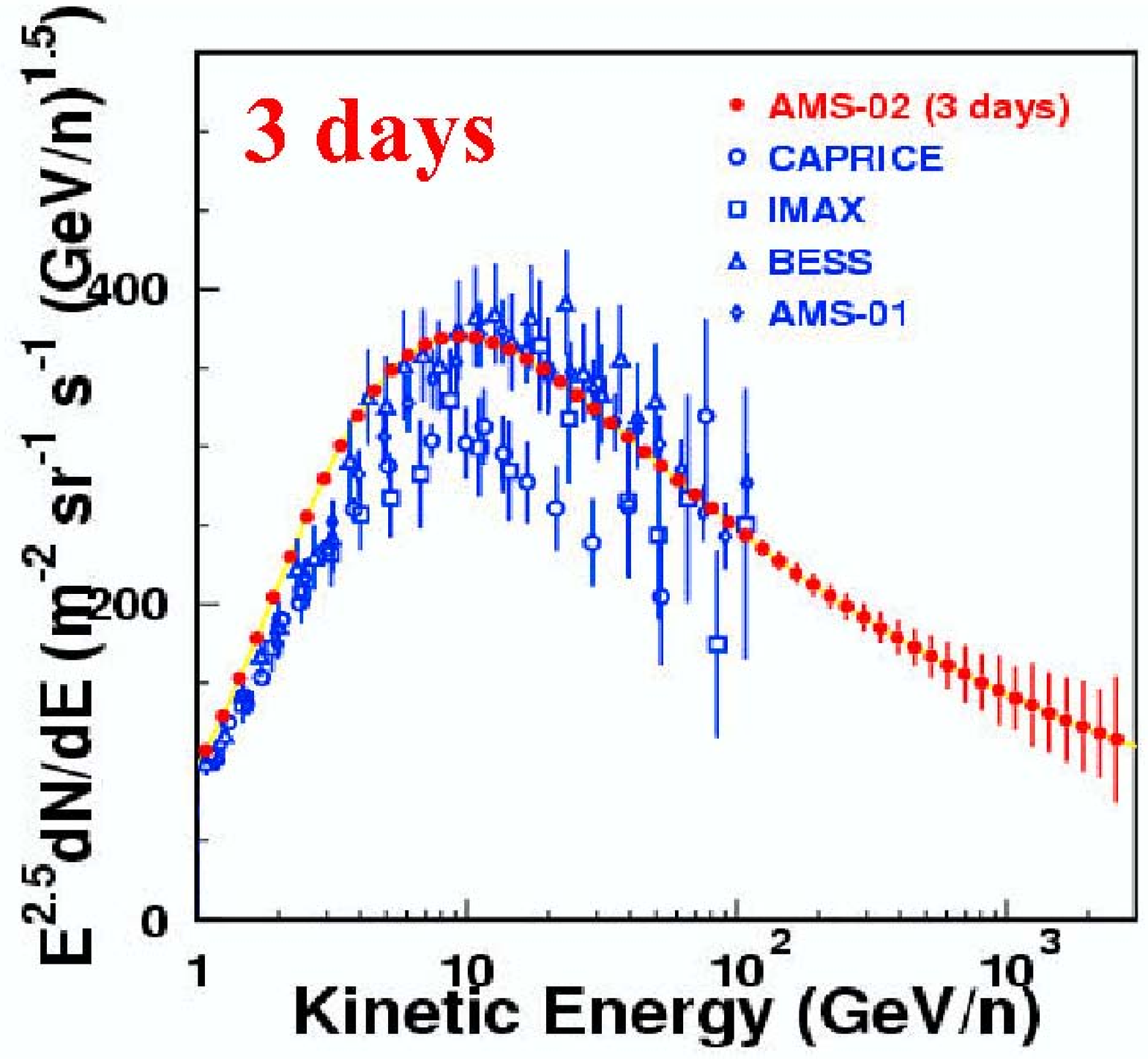}
\includegraphics{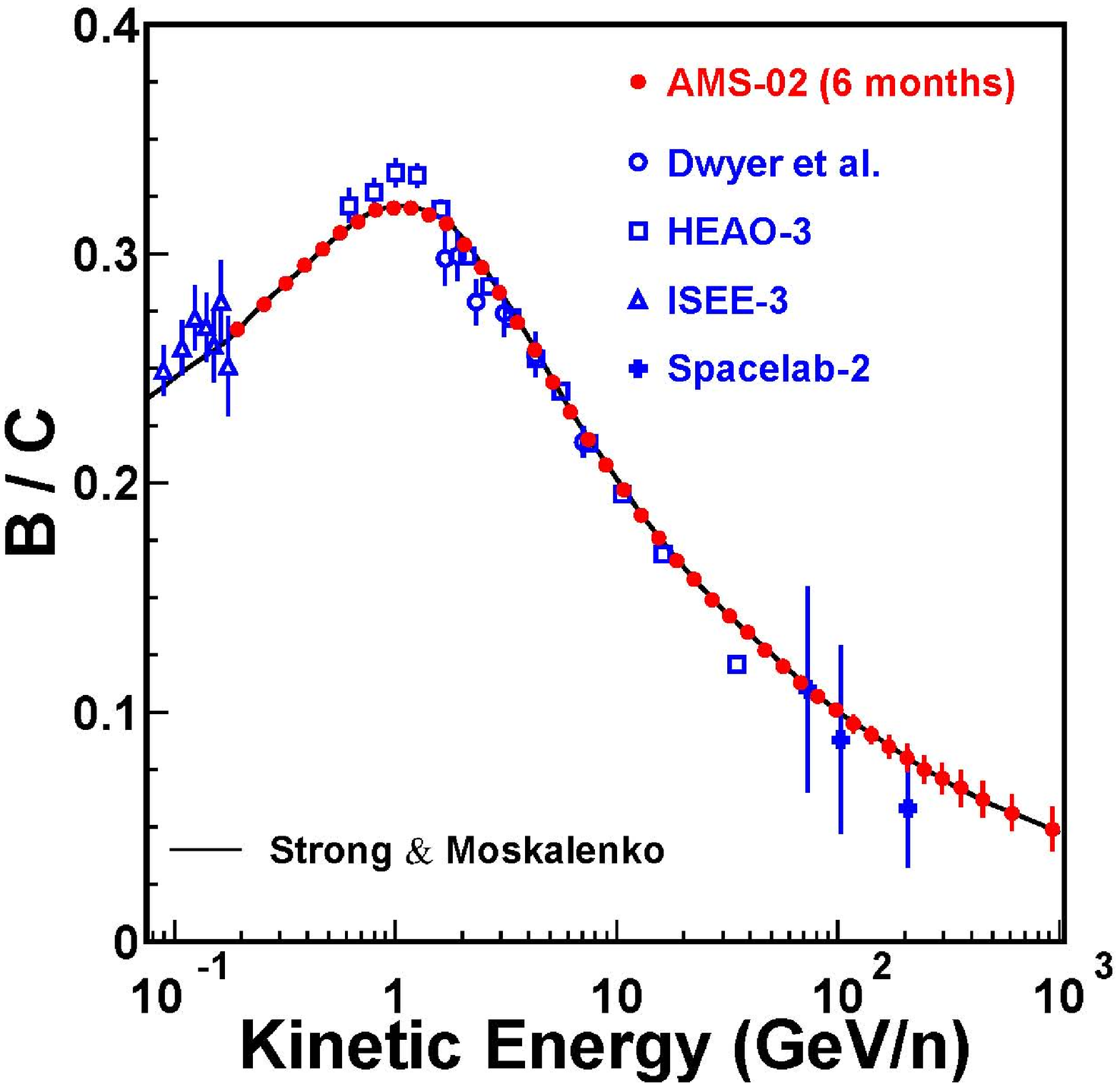}
\includegraphics{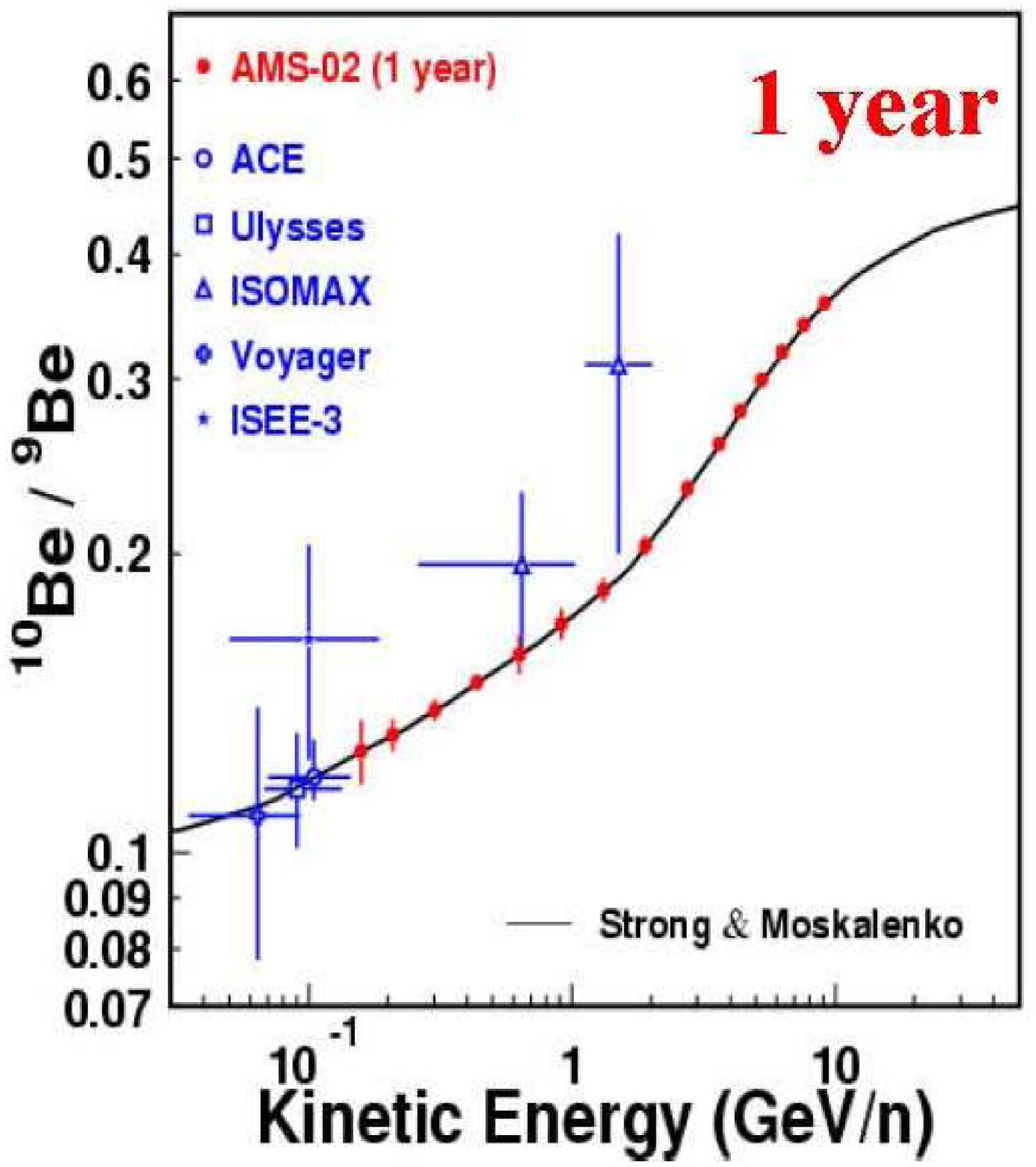} \
      \caption[h]{The present data and the accuracy expected by AMS02 for proton (UL), helium (UR),
      B/C ratio (LL)
      and  $^{10}$Be/$^9$Be  ratio (LR)
      flux measurements.}
     \label{fig4}
    \end{figure}

\section{Conclusions}
The AMS02 has been designed to measure with ppb accuracy primary
cosmic ray composition up to TeV region. These accurate measurements
will allow better understanding of propagation and confinement
mechanisms in our galaxy. The study of rare components will allow to
search of new phenomena (dark matter, strangelets (Madsen \& Larsen,
2003)) or to better constrain the fundamental issues as the
existence of primordial antimatter.
\section{Acknowledgements}
I wish to thank the organizers for their kind invitation to this
stimulating and well organized Workshop.

\end{document}